\documentclass[12pt]{article}
\usepackage[dvips]{graphicx}
\usepackage{amsmath}
\renewcommand{\baselinestretch}{1.20}
\begin{document}

hep-ph/0701057
\begin{flushright}
OU-HET-572, \, January, 2007 \ \ \\
\end{flushright}
\vspace{0mm}
\begin{center}
\large{On the D-term of the nucleon generalized parton distributions}
\end{center}
\vspace{0mm}
\begin{center}
M.~Wakamatsu\footnote{Email \ : \ wakamatu@phys.sci.osaka-u.ac.jp}
\end{center}
\vspace{-4mm}
\begin{center}
Department of Physics, Faculty of Science, \\
Osaka University, \\
Toyonaka, Osaka 560-0043, JAPAN
\end{center}

PACS numbers : 13.60.Hb, 12.39.Fe, 12.39.Ki, 14.20.Dh

\vspace{6mm}
\begin{center}
\small{{\bf Abstract}}
\end{center}
\vspace{-2mm}
\begin{center}
\begin{minipage}{15.5cm}
\renewcommand{\baselinestretch}{1.0}
\small
It is known that some of the deeply-virtual-Compton-scattering
observables, for instance, the beam-charge asymmetry in the hard
electroproduction of real photons on the nucleon, are extremely
sensitive to the magnitude of D-term appearing in the parameterization
of the generalized parton distributions. We report a theoretical
analysis of both the isoscalar and isovector parts of the nucleon
D-term within the framework of the chiral quark soliton model,
without recourse to the derivative expansion type approximation
used in previous works.

\normalsize
\end{minipage}
\end{center}
\renewcommand{\baselinestretch}{2.0}

\vspace{8mm}

Growing attention has recently been paid to the studies of the
so-called generalized parton distributions (GPDs) not only because
they provide us with the unique means to experimentally access the
quark orbital angular momentum in the nucleon but also because
they offer the most detailed information on the underlying
quark-gluon structure of the nucleon \cite{Mueller94}
\nocite{Ji97}-\nocite{Ji98}\nocite{Rad01}\nocite{GPV01}
\nocite{Diehl03}\cite{BR05}.
However, since the GPDs are functions of three kinematical
variables and since they appear as complicated convolution
integrals in the cross section formulas of the deeply-virtual
Compton scatterings (DVCS), deeply-virtual meson productions
(DVMP) etc., a suitable parameterization of them is practically
unavoidable.
The most popular parameterization of GPDs is to use the
double distributions \cite{Mueller94},\cite{Rad01} supplemented
with the so-called D-term \cite{PW99}.
It turned out that some of the DVCS observables, for instance,
the beam-charge asymmetry in the hard electroproductions of real
photons on the nucleon, are extremely sensitive to the magnitude
of the D-term necessary in the above parameterization
\cite{GPV01},\cite{Ellinghaus02}.

So far, there has been only a limited number of theoretical
studies on the nucleon D-term. The first estimate of the D-term is
based on the chiral quark soliton model (CQSM), more precisely,
on the CQSM predictions for the GPD $H^{u+d} (x,\xi,t)$ given by
Petrov et al. under the derivative-expansion type
approximation \cite{Petrov98}.
Using these predictions, the authors of \cite{GPV01}
as well as of \cite{Kivel01} estimated several Mellin moments of
$H^{u+d} (x,\xi,t)$ at various values of $\xi$ and $t$ and
extrapolated them to $t=0$.
The results are then fitted to the Gegenbauer expansion given as
\begin{equation}
 \sum_{q = u,d,s, \cdots} \,D^q (z) \ = \ 
 (1 - z^2) \,\sum_{n=1, odd}^\infty \,d_n \,C_n^{3/2} (z).
\end{equation}
This led them to the following estimate for the expansion
coefficients :
\begin{equation}
 d_1 \ \simeq \ - \,4.0, \ \ \ d_3 \ \simeq - \,1.2,
 \ \ \ d_3 \ \simeq \ - \,0.4, \label{d1d3d5}
\end{equation}
with higher coefficients being small. 
We recall here that the $D^q (z)$ in the flavor-singlet channel mixes
under evolution with the corresponding gluon D-term $D^g (z)$.
The numbers quoted in Eq.(\ref{d1d3d5}) corresponds to the values at
a few GeV scale. At the model energy scale around $600 \,\mbox{MeV}$,
their result corresponds to (see the footnote of \cite{Schweitzer02})
\begin{equation}
 d_1 \ \simeq \ - \,8.0 .
\end{equation}
Although used in many recent phenomenological analyses of the DVCS and
the DVMP processes, such an estimate of the D-term coefficients is of
highly qualitative nature.
A little more direct estimate of $d_1$, the first
coefficient of the Gegenbauer expansion, was made by Schweitzer
et al. within the same model \cite{Schweitzer02}.
By starting with the model expression for the unpolarized
GPD $H (x,\xi,t)$, they derived a closed formula
for $d_1$. They also estimate its numerical value by using the
derivative-expansion type approximation to find that
\begin{equation}
 d_1 \ \simeq \ - \,9.46,
\end{equation}
at the model energy scale. After a simple estimate of the scale
dependence, the author of \cite{Schweitzer02} got the number
\begin{equation}
 d_1 \ \simeq \ - \,4.7 \ \ \ \mbox{at a few $\mbox{GeV}^2$} ,
\end{equation}
which they claim is qualitatively consistent with the results of
\cite{Kivel01},\cite{GPV01}.
Although of preliminary nature, there also exists a lattice QCD
study of the generalized form factors $C_2^q (t)$ \cite{QCDSF04},
forward limit of which are related to $d_1$.

Clearly, in spite of its phenomenological importance, we must say
that our knowledge on the precise magnitude of the D-term is still
rather poor and uncertain.
In view of the circumstance above, we think it useful to
evaluate the most important parameter of the D-term, i.e. the first
coefficient of the Gegenbauer expansion of the D-term within the
framework of the CQSM, without recourse to the derivative-expansion
type approximation.
In the present paper, we try to estimate not only the isoscalar
part of $d_1$ but also its isovector part, which is subleading
in the $1 / N_c$ expansion. For the sake of comparison,
we also derive the theoretical expression
of $d_1^{u+d}$ in the familiar MIT bag model, and see what prediction
it gives.

To fix the normalization convention of the relevant quantities,
it would be convenient to start our investigation with the
nucleon matrix elements of the
quark and gluon parts of the (symmetric) QCD energy-momentum tensor
parameterized by four form factors as \cite{Polyakov03},\cite{Ji97}
\begin{eqnarray}
 \langle p^\prime | \,\hat{T}_{\mu \nu}^{q,g} (0) \,| p \rangle
 &=& \bar{N} (p^\prime) \,\left[ \,M_2^{q,g} (t) \,
 \frac{P_\mu \,P_\nu}{M_N} \ + \ J^{q,g} (t) \,
 \frac{i \,P_{\{\mu}\,\sigma_{\nu\}\rho} \,\Delta^\rho}{M_N}
 \right. \nonumber \\
 &+& \left. d^{q,g} (t) \,\frac{1}{5 \,M_N} \,
 \left( \Delta_\mu \,\Delta_\nu - g_{\mu \nu} \,\Delta^2 \right)
 \ + \ \bar{c}^{q,g} (t) \,g_{\mu \nu} \,\right] \,N(p) ,
 \label{Eq:tmat}
\end{eqnarray}
where $P = (p^\prime + p) / 2, \Delta = p^\prime - p$, and
$t = \Delta^2$.
Here $\hat{T}^q_{\mu \nu} = \bar{q} \,
\gamma_{\{\mu} \overleftrightarrow{\nabla}_{\nu\}} q$
is the QCD energy-momentum tensor of the quark with flavor $q$,
while $\hat{T}_{\mu \nu}^g = G_{\mu \alpha}^a \,G_{\alpha \nu}^a
+ \frac{1}{4} \,G^2$ is the corresponding gluon part.
(The form factor $\bar{c}(t)$ accounts for nonconservation
of the separate quark and gluon parts of the energy-momentum
tensor. They must satisfy the constraint 
$\sum_q \,\bar{c}^q (t) + c^g (t) = 0$ due to the conservation
of the total (quark plus gluon) energy-momentum tensor.)
The form factors in Eq.(\ref{Eq:tmat}) are related to the 2nd Mellin
moments of the familiar unpolarized GPDs
$H^q (x,\xi,t)$ and $E (x,\xi,t)$ as
\begin{eqnarray}
 \sum_q \,\int_{-1}^1 \,x \,H^q (x,\xi,t) \,dx &=&
 M_2^Q (t) \ + \ \frac{4}{5} \,d^Q (t) \,\xi^2, \\
 \sum_q \,\int_{-1}^1 \,x \,E^q (x,\xi,t) \,dx &=&
 2 \,J^Q (t) \ - \ M_2^Q (t) \ - \ \frac{4}{5} \,d^Q (t) \,\xi^2.
\end{eqnarray}
Here, the suffix $Q$ denotes the summation over all quark
flavors, for example, $J^Q (t) \equiv \sum_{q = u,d,s, \cdots} \,
J^q (t)$. (Practically, we confine here to the light-quark
components of two flavors, which means that $Q = u + d$.)
The sum of the above two equations with $t = 0$ gives the famous
Ji's angular momentum sum rule \cite{Ji97},\cite{Ji98} :
\begin{equation}
 \sum_q \,\int_{-1}^1 \,x \,\left(\, H^q (x,\xi,0) \ + \ 
 E^q (x,\xi,0) \,\right) \,dx \ = \ 2 \,J^Q,
\end{equation}
with $J^Q$ being the total angular momentum carried by the quark
fields in the nucleon.

The interest of our present study is the forward limit of $d^Q(t)$,
which just corresponds to the first coefficients in the
Gegenbauer expansion of the so-called D-term, i.e.
\begin{equation}
 d_1^{u+d} \ \equiv \ d^Q (0).
\end{equation}
According to Polyakov \cite{Polyakov03}, the constants $d_1^{u+d}$
can be expressed as
\begin{equation}
 d_1^{u+d} \ = \ - \,\frac{M_n}{2} \,\int d^3 r \,
 T^Q_{ij} (\mbox{\boldmath $r$}) \,\left( \,
 r^i \,r^j \ - \ \frac{1}{3} \,\delta^{ij} \,r^2 \,\right).
\end{equation}
Here $T^Q_{\mu \nu} (\mbox{\boldmath $r$})$ is the net quark
contribution to the static energy-momentum tensor density of
the nucleon defined by
\begin{equation}
 T^Q_{\mu \nu} (\mbox{\boldmath $r$}) \ = \ \frac{1}{2 \,M_N} \,
 \int \,\frac{d^3 \Delta}{(2 \,\pi)^3} \,
 e^{- \,i \,\mbox{\boldmath $\Delta$} \cdot \mbox{\boldmath $r$}}
 \langle p^\prime, S | \,\hat{T}^Q_{\mu \nu} (0) \,| p, S \rangle ,
\end{equation}
with $\hat{T}^Q_{\mu \nu} (0)$ the quark part of the QCD energy-momentum
tensor operator.
As expected, the D-term is seen to contain valuable
information on the distribution of energy-momentum tensor inside the
nucleon.

As a warm-up, let us first evaluate $d_1$ in a simple model of baryons,
i.e. the MIT bag model. We must calculate
\begin{equation}
 d_1^{u+d} ({\rm MIT}) \ = \ 
 - \,\frac{M_N}{2} \,\langle \Psi_{gs} \,| \,
 r^2 \,[ \,\mbox{\boldmath $\alpha$} \cdot \hat{\mbox{\boldmath $r$}}
 \,\overleftrightarrow{\mbox{\boldmath $p$}} \cdot \hat{\mbox{\boldmath $r$}}
 \ - \ \frac{1}{3} \,\mbox{\boldmath $\alpha$} \cdot
 \overleftrightarrow{\mbox{\boldmath $p$}} \,] \,
 | \,\Psi_{gs} \rangle ,
\end{equation}
with $\alpha^i = \gamma^0 \gamma^i$ the standard Dirac matrices. Here
\begin{equation}
 \Psi_{gs} (\mbox{\boldmath $r$}) \ = \ N \,
 \left( \begin{array}{c}
 j_0 (kr) \,| \,(l=0) \,j = 1/2 , m \rangle \\
 - \,i \,j_1 (kr) \,| \,(l=1) \,j = 1/2, m \rangle \\
 \end{array} \right) ,
\end{equation}
with
\begin{equation}
 N^{-2} \ = \ 2 \,R^2 \,(kR - 1) \,j_0 (kR), \ \ \ 
 \omega_0 \ \equiv \ k R \ \simeq \ 2.043
\end{equation}
is the ground state wave function with $R$ being the bag radius.
After some manipulation, we easily find that
\begin{equation}
 d_1^{u+d} ({\rm MIT}) \ = \ - \,M_N \,N^2 \,k \,
 \left\{ \, \,\int_0^R \,j_1 (kr) \,r^4 \,j_1 (kr) \,dr \ - \ 
 \int_0^R \,j_0 (kr) \,r^4 \,j_2 (kr) \,dr \,\right\} . \label{Eq:d1MIT}
\end{equation}
For an order of magnitude estimate for $d_1^{u+d} ({\rm MIT})$,
here we use the bag model parameter adopted by Jaffe and Ji
\cite{Jaffe92}, i.e. $M_N R \simeq 4.0 \,\omega_0$.
Eq.(\ref{Eq:d1MIT}) then gives
\begin{equation}
 d_1^{u+d} ({\rm MIT}) \ \simeq \ - \,0.716,
\end{equation}
which turns out to be many times smaller in magnitude than the previous
estimates based on the CQSM, $d_1^{u+d} \simeq
- \,(8.0 \sim 9.5)$.

Next, we derive the theoretical expression for $d_1$ within the
CQSM. We first note that the energy-momentum tensor in the CQSM
formally takes the same form as that of QCD, since the effective
pion degrees of freedom contained in it is not an independent
fields of quarks. The leading order contribution to the isoscalar
$d_1$ comes from the zeroth order term in the collective rotational
velocity of the soliton \cite{DPP88},\cite{WY91}, so that it is
given in the following form :
\begin{equation}
 d_1^{u+d} \ = \ - \,\frac{M_N}{2} \,N_c \,\,\sum_{n \leq 0} \,\,
 \langle n \,| \,
 r^2 \,[ \,\mbox{\boldmath $\alpha$} \cdot \hat{\mbox{\boldmath $r$}}
 \,\overleftrightarrow{\mbox{\boldmath $p$}} \cdot \hat{\mbox{\boldmath $r$}}
 \ - \ \frac{1}{3} \,\mbox{\boldmath $\alpha$} \cdot
 \overleftrightarrow{\mbox{\boldmath $p$}} \,] \,
 | \,n \rangle .
\end{equation}
The symbol $\sum_{n \leq 0}$ denotes the summation over the occupied
states (i.e., the discrete valence level ($n = 0$) plus the
negative-energy Dirac-sea orbitals ($n < 0$)) in the hedgehog
mean field.
In the above equation, $\overleftrightarrow{\mbox{\boldmath $p$}}
= \frac{1}{2} \,( \overrightarrow{\mbox{\boldmath $p$}} + 
\overleftarrow{\mbox{\boldmath $p$}} )$ with
$\overrightarrow{\mbox{\boldmath $p$}}$ and
$\overleftarrow{\mbox{\boldmath $p$}}$ being the momentum operators
respectively acting on the initial and final state wave functions.
After some manipulation,
$d_1^{u+d}$ can be transformed into a form, which is convenient
for numerical calculation : 
\begin{equation}
 d_1^{u+d} \ = \ 
 - \,\frac{\sqrt{4 \,\pi}}{\sqrt{6}} \,M_N \,N_c \,
 \sum_{n \leq 0} \,\,\langle n \,| \,r^2 \,\left[ \,
 [ Y_2 (\hat{\mbox{\boldmath $r$}}) \times 
 \overleftrightarrow{\mbox{\boldmath $p$}}]^{(1)} \,\times
 \mbox{\boldmath $\alpha$} \,\right]^{(0)} \,| \,n \rangle .
 \label{d1theory1}
\end{equation}
It is an easy exercise that this precisely coincides with the
following expression
\begin{equation}
 d_1^{u+d} \ = \ - \,\frac{5}{4} \,N_c \,M_N \, \sum_{n \leq 0} \,
 \langle n \,| \,\gamma^0 \,\gamma^3 \,\left\{ \,
 \overleftrightarrow{p}^3, | \mbox{\boldmath $x$} |^2 \,
 P_2 (\cos \theta) \,\right\} \,| n \rangle , \label{d1theory2}
\end{equation}
derived by Schweitzer et al. in the same model by starting
with the expression for the nucleon unpolarized GPD
$H^{u+d} (x,\xi,t)$. 
First pointing out that the valence quark contribution to $d_1^{u+d}$
vanishes identically, i.e. $d_1^{u+d} ({\rm val}) = 0$, they
estimated the contribution of the polarized Dirac sea by means
of a kind of derivative-expansion type approximation.
They thus find that $d_1^{u+d} = d_1^{u+d} ({\rm sea}) \simeq - \,9.46$
at the model energy scale around $600 \,\mbox{MeV}$.

However, we find no reason why $d_1^{u+d} ({\rm val})$ vanishes.
In fact, the relevant operator appearing in Eq.(\ref{d1theory1})
is a positive parity operator with the total grand spin $K$ being
zero (here $\mbox{\boldmath $K$} = \mbox{\boldmath $J$} + 
\frac{1}{2} \,\mbox{\boldmath $\tau$}$),
and there is no selection rule which enforces its matrix element
between the valence quark state with the quantum numbers
$K^P = 0^+$ to vanish. (Although not so obvious, the operator
appearing in Eq.(\ref{d1theory2}) also contains the $K = 0$
component.)
Here, we shall calculate this discrete valence level contribution
explicitly. We also try to evaluate the contribution of the
deformed Dirac sea, without recourse to the derivative expansion type
approximation.
This is possible with use of the discretized momentum basis of
Kahana,Ripka and Soni \cite{KR84},\cite{KRS84}.
In the present analysis, we use the self-consistent
soliton solutions obtained in \cite{WN06} within the double-subtraction
Pauli-Villars regularization scheme \cite{KWW99}.
The model in the chiral limit
contains only one parameter $M$, i.e. the dynamical quark mass.
Here, we use the value $M = 400 \,\mbox{MeV}$. Since the
energy-momentum-tensor distribution carried by the quark fields
is expected to be sensitive to the value of the pion mass, we
shall also investigate the pion mass dependence of $d_1^{u+d}$.
The effective model lagrangian, which incorporates the finite pion
mass effects, is given in \cite{KWW99}.
Self-consistent soliton solutions
are prepared in \cite{WN06} for several values of pion mass.
As pointed out in that paper, favorable physical predictions of the
model are obtained by using the value of $M = 400 \,\mbox{MeV}$
and $m_\pi = 100 \,\mbox{MeV}$, since this set gives a self-consistent
solution close to the phenomenologically successful one obtained
with $M = 375 \,\mbox{MeV}$ and $m_\pi = 0 \,\mbox{MeV}$ in the
single-subtraction Pauli-Villars regularization scheme in the
previous studies of nucleon parton distribution functions
\cite{DPPPW96}\nocite{DPPPW97}\nocite{WGR96}
\nocite{GRW98}-\nocite{WK99}\nocite{Waka03}\cite{Waka06}.
We first show the theoretical prediction for $d_1^{u+d}$ corresponding
to this favorable parameter set. It gives
\begin{equation}
 d_1^{u+d} \ = \ d_1^{u+d} ({\rm val}) \ + \ d_1^{u+d} ({\rm sea}),
\end{equation}
with
\begin{equation}
 d_1^{u+d} ({\rm val}) \ \simeq \ 0.66, \ \ \ 
 d_1^{u+d} ({\rm sea}) \ \simeq \ - 5.51.
\end{equation}
We confirm that the dominant contribution to $d_1^{u+d}$ comes from
the quarks in the negative-energy Dirac-sea orbitals.
This deformed Dirac-sea contribution is large and negative.
However, we also find that the quarks in the discrete valence level
gives nonzero and positive contribution,
which partially cancels the Dirac-sea contribution.
As pointed out in \cite{GPV01}, the large and negative prediction
for the D-term coefficient is a characteristic feature of the CQSM,
which maximally incorporates the spontaneous breaking of the chiral
symmetry. In fact, the dominant contribution from the polarized
Dirac sea is also viewed as simulating the $t$-channel exchange
of two pions with the quantum numbers $J^{PG} = 0^{++}, 2^{++}, \cdots$
\cite{Diehl03}.
The final model predictions for $d_1^{u+d}$ obtained as the sum
of the valence and the Dirac-sea contribution is
\begin{equation}
 d_1^{u+d} \ \simeq \ - 4.85,
\end{equation}
which is a little smaller than the prediction $d_1^{u+d} \simeq -8.0$
obtained from the numerically evaluated $H^{u+d} (x,\xi,t)$
\cite{Kivel01},\cite{Petrov98} and the
prediction $d_1^{u+d} \simeq - 9.46$ obtained based on the
derivative-expansion type approximation with neglect of the valence
level contribution in the same model \cite{Schweitzer02}.
In view of the difference of the soliton profile functions used
in all these analyses (note that our result corresponds to
$d_1^{u+d} \simeq - \,6.2$ in the chiral limit), the
qualitative agreement is encouraging, and it confirms the unique
feature of the CQSM, which takes account not only of three valence
quarks but also infinitely many Dirac-sea quarks in the mean potential.
This is clear, if one compares the above predictions of the CQSM with
that of the MIT bag model, $d_1^{u+d} ({\rm MIT}) \simeq - 0.716$.
We also recall that the lattice QCD simulation performed by the
QCDSF collaboration in the heavy pion region around
$m_\pi \sim 800 \,\mbox{MeV}$ \cite{QCDSF04} gives fairly small
number : $d_1^{u+d} ({\rm QCDSF}) = \frac{5}{4} \,C_2^{u+d} (0) \simeq
-0.25 \pm 0.13$. Although the lattice QCD prediction quoted here
corresponds to the energy scale around $Q^2 \simeq 4 \,(\mbox{GeV})^2$,
there seems to be more difference in magnitude than
explained by its scale dependence.

\begin{figure}[htb]
\begin{center}
  \includegraphics[height=.36\textheight]{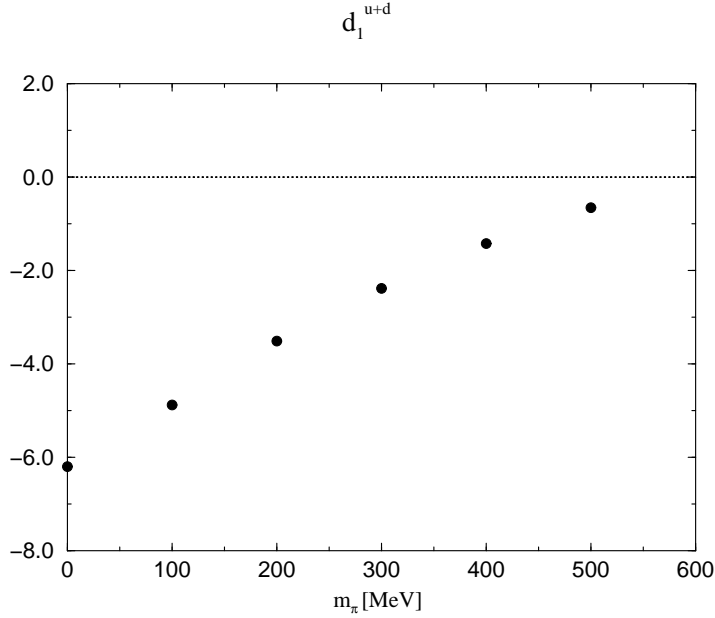}
  \caption{\baselineskip16pt The CQSM prediction for $d_1^{u+d}$
  in dependence of the pion mass.}%
\label{fig1:xg1d}
\end{center}
\end{figure}

The pion cloud interpretation of the Dirac-sea contribution to
$d_1^{u+d}$ in the CQSM may also be confirmed by investigating
its pion mass dependence. We show in Fig.1 the CQSM prediction
for $d_1^{u+d}$ in dependence of the pion mass.
One sees that the magnitude of
$d_1^{u+d}$ rapidly decreases as $m_\pi$ increases, showing a
tendency to match the very small lattice QCD prediction
obtained in the heavy pion region at least qualitatively.
Since the valence quark contribution to $d_1^{u+d}$ is less sensitive
to the variation of the pion mass, it can be interpreted as the
reduction of the pion cloud effects as the model parameter
$m_\pi$ increases.

\begin{figure}[htb]
\begin{center}
  \includegraphics[height=.36\textheight]{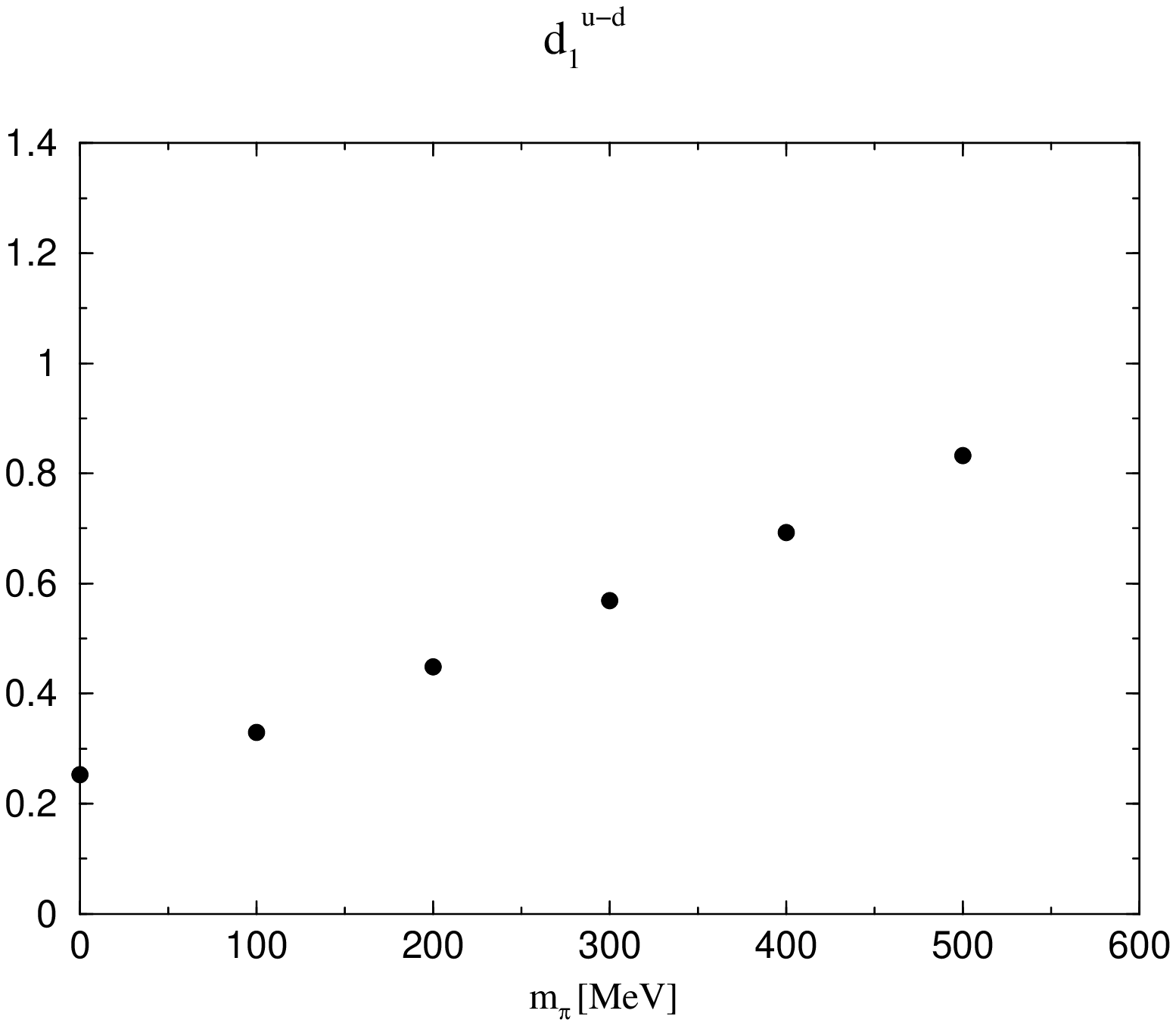}
  \caption{\baselineskip16pt The CQSM prediction for $d_1^{u-d}$
  in dependence of the pion mass.}%
\label{fig1:xg1d}
\end{center}
\end{figure}

Now, we turn to the discussion of the isovector
$d_1$. It was emphasized in \cite{GPV01} that the isovector D-term
is suppressed relative to the isoscalar one by a factor of $1 / N_c$,
so that it is negligible in the large $N_c$ limit.
Since $N_c = 3$ in reality,
however, it is not self-evident whether it is in fact numerically
small or not. Here, we try to evaluate this subleading term in the
$1 / N_c$ expansion in an explicit manner.
In conformity with the observation above,
$d_1^{u-d}$ in the CQSM survives only at the 1st order in the
collective rotational velocity of the soliton, so that the answer
is given in the following form, i.e. as a double
sum over the single quark orbitals in the hedgehog mean field : 
\begin{eqnarray}
 d_1^{u-d} &=& - \,\frac{\sqrt{4 \,\pi}}{\sqrt{6}} \,M_N \,
 \frac{N_c}{6 \,I} \,\sum_{m > 0, n \leq 0} \,
 \frac{1}{E_m - E_n} \,
 \langle m \,|| \,\mbox{\boldmath $\tau$} \,|| \,n \rangle \nonumber \\
 &\,& \hspace{28mm} \times \ 
 \langle m \,|| \,r^2 \,\left[ \,[ \,Y_2 (\hat{\mbox{\boldmath $r$}})
 \times \overleftrightarrow{\mbox{\boldmath $p$}}]^{(1)} \times
 \mbox{\boldmath $\alpha$} \,\right]^{(0)} \,
 \mbox{\boldmath $\tau$} \, || n \rangle .
\end{eqnarray}

Let us first show the prediction obtained with the favorable set of
the model parameters, i.e. $M = 400 \,\mbox{MeV}$ and
$m_\pi = 100 \,\mbox{MeV}$. This gives
\begin{eqnarray}
 d_1^{u-d} &=& d_1^{u-d} ({\rm val}) \ + \ d_1^{u-d} ({\rm sea})
 \ \simeq \ 0.33 \ + \ 0.01 \ \simeq \ 0.34.
\end{eqnarray}
We find that, in contrast to the isoscalar case, the contribution
of the discrete valence level is dominant,
while that of the deformed Dirac-sea is almost negligible.
We also confirm that net $d_1^{u-d}$ is much smaller in magnitude
than $d_1^{u+d}$, in conformity with the expectation based on the
large $N_c$ counting.
Still, our explicit calculation shows that the difference of $d_1^u$
and $d_1^d$ takes nonzero positive value owing to the presence of
the discrete valence level contribution. Also interesting is
how the isovector $d_1$ depends on the pion mass.
We show in Fig.2 the CQSM prediction for $d_1^{u-d}$ in dependence of
the pion mass.
Contrary to the isoscalar case, the magnitude of $d_1^{u-d}$ is an
increasing function of $m_\pi$. This peculiar behavior of
$d_1^{u-d}$ resembles the pion mass dependence of the generalized
form factor $B_{20}^{u-d} (t)$ at $t=0$,
or equivalently the isovector gravito anomalous magnetic
moment of the nucleon, so that it may have a similar
origin. (See Fig.5 of \cite{WN06}, and the explanation around there.)

In summary, we have carried out a theoretical analysis of the
most important constants that characterize the nucleon D-term,
i.e. the first coefficients $d_1$ of its Gegenbauer expansion,
within the framework of the CQSM, without recourse to the
derivative-expansion type approximation used in the previous studies.
We gave predictions not only for the leading isoscalar part of $d_1$
but also for the subleading isovector part of $d_1$ in the $1 / N_c$
expansion.
Our treatment makes it possible to estimate the contribution of
the discrete valence level and that of the negative-energy
Dirac-sea levels separately. We found that, as for the isoscalar
$d_1^{u+d}$, the contribution of the deformed Dirac-sea is
large and negative and dominates over the small positive contribution
from the discrete valence level. This reconfirms the unique feature
of the CQSM, which takes good account of the effects of the pion cloud
generated by the spontaneous chiral symmetry breaking of the QCD vacuum.
The predicted value of $d_1^{u+d} \simeq - \,(4.9 \sim 6.2)$ at the model
energy scale around $600 \,\mbox{MeV}$ qualitatively supports
the previous estimate $d_1^{u+d} \simeq - \,(8.0 \sim 9.6)$ obtained in
the same model based on the derivative-expansion type approximation.
We have also found that $d_1^{u-d} \simeq 0.34$, which we confirm is
much smaller than $d_1^{u+d}$ but cannot be completely neglected.
We hope that the theoretical analysis carried out here will
give useful constraints on the future analyses of high-energy DVCS
and DVMP processes using the double distribution parameterization
of the nucleon GPDs supplemented with the D-term.

\vspace{10mm}
\noindent
\begin{large}
{\bf Acknowledgement}
\end{large}

\vspace{3mm}
This work is supported in part by a Grant-in-Aid for Scientific
Research for Ministry of Education, Culture, Sports, Science
and Technology, Japan (No.~C-16540253)

\vspace{10mm}
\noindent
\begin{large}
{\bf Note added}
\end{large}

\vspace{3mm}
After submission of the present paper, two papers \cite{GGOSSU2007A}
and \cite{GGOSSU2007B} appeared where the constant $d_1^{u+d}$
(in addition to some other nucleon form factors of the energy momentum tensor)
was calculated in a different manner within the same model with
different regularization scheme. In the footnote 4 of \cite{GGOSSU2007A},
it was pointed out that the statement $d_1^{u+d} (val) = 0$ made in
\cite{Schweitzer02} is incorrect, which is consistent with our
observation in the present paper.

%
%

\setlength{\baselineskip}{5mm}

\end{document}